\documentclass[prl,aps,preprint]{revtex4}

\begin{document} 

\title{Slow oscillations of magnetoresistance in quasi-two-dimensional 
metals} 
\author{M. V. Kartsovnik$^{1}$, P. D. Grigoriev$^{2,3}$, 
W. Biberacher$^{1}$,  N. D. Kushch$^4$ and P. Wyder$^2$ } 
\affiliation{$^{1}$ Walther-Mei\ss ner-Institut, Bayerische Akademie der 
Wissenschaften, Walther-Mei\ss ner-Str. 8, D-85748 Garching, Germany\\ 
$^{2}$ Grenoble High Magnetic Field Laboratory, MPI-FKF and CNRS, 
BP 166, F-38042 Grenoble Cedex 09, France \\ 
$^{3}$ L. D. Landau Institute for Theoretical Physics, 142432 Chernogolovka, 
Russia \\ 
$^4$ Institute of Problems of Chemical Physics, 142432 Chernogolovka, 
Russia } 

\date{\today } 

\begin{abstract} 
Slow oscillations of the interlayer magnetoresistance observed in the 
layered organic metal $\beta $-(BEDT-TTF)$_2$IBr$_2$ are shown to originate 
from the slight warping of its Fermi surface rather than from independent 
small cyclotron orbits. Unlike the usual Shubnikov-de Haas effect, these 
oscillations are not affected by the temperature smearing of the Fermi 
distribution and can therefore become dominant at high enough temperatures.  
We suggest that the slow oscillations are a general feature of clean 
quasi-two-dimensional metals and discuss possible applications of the 
phenomenon. 
\end{abstract} 

\maketitle

\newpage 
Quantum oscillations of magnetization (de Haas-van Alphen, dHvA, effect) and 
magnetoresistance (Shubnikov-de Haas, SdH, effect) in strong magnetic fields 
have been successfully used for many years in order to investigate Fermi 
surfaces and other electronic properties of metals \cite{Shoe}. In 
three-dimensional (3D) metals the oscillations are determined by only a small 
fraction of conduction electrons near extremal cyclotron orbits on the Fermi 
surface (FS). Therefore they are very weak and can perfectly be described in 
terms of the so-called Lifshitz-Kosevich (LK) formalism \cite{Shoe}. 
Recently, however, materials of lower dimensionality such as two-dimensional 
electron gas systems, layered metal oxides and organic conductors have 
been of high interest. In these materials the relative number of electrons 
contributing to the oscillations is much higher than in 3D metals and the 
main assumptions of the standard theory, i.e. a weak oscillation amplitude 
and constant chemical potential, are often no longer fulfilled. 

Some crystalline organic metals, being very clean and having extremely large 
anisotropy of electronic properties, are excellent objects for studying 
specific features of the magnetic quantum oscillations in the low-dimensional 
limit \cite{Jero,Harr,Wosn1,Balt,PhSh}.  Various deviations of the 
oscillatory magnetization from the LK theory observed in these compounds 
can be fairly well explained by a semi-phenomenological description of the 
quasi-two-dimensional (Q2D) dHvA effect developed in a number of theoretical 
works \cite{Harr,Gv1,dHvADiff,Pavel2}. The situation with the SdH effect is 
less satisfactory. Despite some progress in understanding several anomalies 
of the oscillatory magnetoresistance \cite{Harr,PhSh,ShubDiff,Av} a 
consistent theory is still lacking. Moreover, there are still open 
qualitative questions. 
One of these questions concerns the origin of slow oscillations of 
magnetoresistance which were first found in the Q2D organic metal $\beta 
$-(BEDT-TTF)$_2$IBr$_2$ \cite{Kar1,Kar2}. Similar oscillations have been 
observed in other layered organic conductors, e.g. 
$\beta $-(BEDT-TTF)$_2$I$_3$ \cite{Kar3,Wosn2}, 
$\kappa $-(BEDT-TTF)$_2$Cu$_2$(CN)$_3$ \cite{Ohmi}, 
and $\kappa $-(BEDT-TSF)$_2$C(CN)$_3$ \cite{togo}. Since the 
behavior of these oscillations strongly resembles that of the SdH effect, 
they have been supposed to originate from additional very small pockets of 
the FS. However, the band structure calculations which are basically 
believed to correctly reproduce the FS topology of organic metals (see e.g. 
\cite{OMRev} for a review) do not explain such small groups of carriers in 
any of the cited compounds. 

In this Letter, we report on detailed studies of the oscillating interlayer 
magnetoresistance and magnetization of $\beta $-(BEDT-TTF)$_2$IBr$_2$ at 
various orientations of magnetic field. Our results provide an 
unequivocal evidence that the slow oscillations of the magnetoresistance do 
not reveal any new carriers but are ultimately caused by a weak warping of 
the single cylindrical FS in this Q2D metal. We propose a theoretical 
explanation of the phenomenon which appears to be in good agreement with the 
experiment. 

The experiment was performed on a high-quality 
$[R(290 {\text K})/R(2 {\text K}) \geq 3000$] 
single crystal with the dimensions $0.6\times 0.3\times 0.12 {\text {mm}}^3$. 
The sample was mounted into a measuring cell in a $^3$He-cryostat allowing 
simultaneous registration of the resistance and magnetic torque \cite{Weis} 
at different orientations of the magnetic field produced by a 14~T 
superconducting magnet. The field orientation was defined by the angle 
$\theta$ between the field direction and the normal to the highly conducting 
$ab$-plane. 

The FS of $\beta $-(BEDT-TTF)$_2$IBr$_2$ is a slightly warped cylinder with 
the axis along the least conducting direction \cite{Kar2,Wosn3,Kar4}. The 
general behavior of the oscillating interlayer 
magnetoresistance and magnetization is shown in Fig.~1 and fully agrees with 
the previous reports \cite{Kar2,Wosn3}. 
Cyclotron orbits on the cylindrical FS give rise to rapid SdH (Fig.~1a) and 
dHvA (Fig.~1b) oscillations with the frequency $F = 3930$~T that corresponds 
to the FS cross-sectional area $S\approx 0.53S_{\text BZ}$ 
($S_{\text BZ}$ is the Brillouin zone area). The amplitudes of both the SdH 
and dHvA oscillations are clearly modulated due to a slight warping of the 
FS. A comparison of the beat and fundamental frequencies yields the 
evaluation of the warping: $\Delta S/S=2F_{\text {beat}}/F \simeq 10^{-2}$. 

In addition to the rapid SdH oscillations, the magnetoresistance exhibits 
prominent slow oscillations with the frequency $F_{\text {{slow}}}\approx 
42$ T. Due to a large cyclotron mass, $m\approx 4.2m_e$ ($m_e$ is the free 
electron mass), the fundamental oscillations rapidly diminish with increasing 
temperature and can barely be resolved at the highest field at $T=1.4$~K. By 
contrast, the amplitude of the slow oscillations remains almost the same as 
at 0.6 K. Noteworthy, no trace of slow oscillations has been found 
in our magnetization measurements. 

A clue for understanding the origin of the slow oscillations is the 
dependence of their frequency on the tilt angle $\theta $ displayed in 
Fig.~2a \cite{comment0}. Unlike the rapid oscillations having the 
$1/\cos \theta$-dependence typical of Q2D metals, the slow oscillations show 
a strong non-monotonic change of their frequency with $\theta$. Such a 
behavior immediately reminds that of the beat frequency 
$F_{\text {beat}}(\theta)$. 
At certain magnetic field orientations, repeated periodically with 
$\tan \theta$, the areas of cyclotron orbits in a Q2D metal become 
independent of the orbit positions in $k$-space \cite{Yam}. This obviously 
results in periodic drops of $F_{\text {beat}}$ to zero \cite{Wosn3}. 
At the same angles the semiclassical part of the magnetoresistance turns out 
to sharply increase \cite{AMRO}, giving rise to the so-called angle-dependent 
magnetoresistance oscillations (AMRO) \cite{Kar2,Kar4}. Fig. 2b shows the 
angular dependence of the background magnetoresistance of our sample at 
$B=14$ T, revealing two prominent AMRO peaks at $\theta \approx 33^{\circ}$ 
and $-20^{\circ}$. It clearly correlates with the angular dependence of 
$F_{\text {slow}}$: the latter rapidly decreases as the magnetoresistance 
approaches the AMRO peaks. 

It is of course very tempting to directly compare $F_{\text {{slow}}}$ and 
$F_{\text {beat}}$. Unfortunately, the rapid oscillations were generally 
observed in a relatively small field interval in our experiment, so that the 
beat period could not be reliably measured in a sufficiently wide 
angular range. Nevertheless, estimations made at a few angles (see e.g. the 
data in Fig. 1) reveal the relationship $F_{\text {{slow}}}(\theta) \approx 
2F_{\text {{beat}}}(\theta)$ within the experimental error bar. 
Thus, one can conclude that the slow oscillations and beats of the rapid 
oscillations have the same physical origin, i.e. both are directly related to 
the warping of the cylindrical FS. 

In order to clarify the mechanism responsible for the slow oscillations, we 
first note that the interlayer conductivity contains several factors which, 
in general, oscillate in magnetic field with the frequency $F$ determined by 
the cross-sectional area of the FS cylinder. The amplitudes of the 
oscillations are modulated, due to the warping of the cylinder, with the 
frequency $F_{\text {beat}}=(2t_{\perp}/\epsilon_F)F\ll F$ ($t_{\perp}$ 
is the interlayer transfer integral and $\epsilon_F$ the Fermi energy). 
The product of two oscillating quantities with modulated amplitudes 
$\tilde{\alpha}$ and $\tilde{\beta}$ yields a slowly oscillating term, e.g. 
$(1+\tilde{\alpha} \cos x)(1+\tilde{\beta}\cos x) = 1+(\tilde{\alpha} 
+\tilde{\beta})\cos x + (\tilde{\alpha}\tilde{\beta}/2)\cos 2x 
+\tilde{\alpha}\tilde{\beta}/2$. Here, the last term describes slow 
oscillations. 

To be more explicit, we consider the interlayer conductivity as determined 
from the Boltzmann transport equation \cite{Mah}: 
\begin{equation} 
\sigma _{zz}=e^{2}\int d\epsilon \,\left( -n_{F}^{\prime }\left( \epsilon 
\right) \right) \,I(\epsilon )\tau (\epsilon ),  \label{Sn1} 
\end{equation} 
where $n_{F}^{\prime }\left( \epsilon \right) =-1/\{4T\cosh ^{2}\left[ 
(\epsilon -\mu (B))/2T\right] \}$ is the derivative of the Fermi 
distribution function, $I(\epsilon )\equiv \sum \left| v_{z}(\epsilon 
)\right| ^{2}$ the square of the electron interlayer velocity $v_z$ 
summed over all states at the energy $\epsilon ,$ and $\tau (\epsilon )$ is 
the momentum relaxation time at the energy $\epsilon $. In Born 
approximation, $\tau (\epsilon )$ is inversely proportional to the density 
of states (DoS): 
$\tau (\epsilon )\propto \rho^{-1}(\epsilon )$ and oscillates in a magnetic 
field \cite {Gv1}: 
\begin{equation} 
\tau (\epsilon )\propto \left[ 1+2\sum_{p=1}^{\infty }(-1)^{p}\cos \left( 
\frac{2\pi p\epsilon }{\hbar \omega _{c}}\right) J_{0}\left( \frac{4\pi 
pt_{\perp}}{\hbar \omega _{c}}\right) R_{D}\right] ^{-1},  \label{DoS1} 
\end{equation} 
where $\omega_c=eB/m$ is the cyclotron frequency, and $R_D=\exp(-2\pi 
^2pk_BT_D/\hbar\omega_c)$ is the usual scattering damping factor 
\cite{Shoe} ($T_D$ is the Dingle temperature). 
If the cyclotron energy is comparable to the interlayer transfer integral, 
the oscillations of the quantity $I(\epsilon )$ become also important. They 
are given by \cite{PhSh} 
\begin{equation} 
I(\epsilon )\propto 1+\frac{\hbar \omega _{c}}{\pi t_{\perp}}\sum_{p= 
1}^{\infty } \frac{\left( -1\right) ^{p}}{p}\cos \left( \frac{2\pi 
p\epsilon }{\hbar \omega _{c}}\right) J_{1}\left( \frac{4\pi pt_{\perp 
}}{\hbar \omega _{c}}\right) R_{D}. \label{I1} 
\end{equation} 

The beats of the oscillations of $\tau$ and $I$ are given by the 0-th and 
1-st order Bessel functions $J_0$ and $J_1$ in Eqs. (\ref{DoS1}) and 
(\ref{I1}), respectively. Their product eventually gives rise to a slowly 
oscillating term in the conductivity. Since the FS warping is large enough in 
our case, $4\pi t_{\perp}>\hbar\omega_c$, one can approximate the Bessel 
functions as 
$J_0(z) = \sqrt{2/\pi z}\cos(z-\pi/4)$ and $J_1(z) = 
\sqrt{2/\pi z}\sin(z - \pi/4)$. 
Further, taking into account the weak amplitude of the SdH effect 
(see Fig.~1) and strong harmonic damping (the second harmonic never exceeded 
1\% of the fundamental one in our experiment), we neglect oscillations of 
the chemical potential $\mu$ and restrict our consideration to the lowest 
order in the damping factors. Then, substituting Eqs. (\ref{DoS1}) and 
(\ref{I1}) into Eq. (\ref{Sn1}) and performing the integration over energy at 
finite temperature, we obtain: 
\begin{eqnarray} 
\sigma _{zz} &=&\sigma _{0}\Bigg\{1+2\sqrt{\frac{\hbar \omega _{c}}{2 
\pi^{2}t}}\sqrt{1+a^{2}}\times  \label{Sigf} \\ 
&&\times \cos \left( \frac{2\pi \,\mu }{\hbar \omega _{c}}\right) 
\cos\left( \frac{4\pi t_{\perp}}{\hbar \omega _{c}}-\frac{\pi }{4}+\phi 
\right) R_{D}R_{T}+  \nonumber \\ 
&&+\frac{\hbar \omega _{c}}{2\pi ^{2}t_{\perp}}\left[ 1+\sqrt{1+a^{2}}\cos 
\left( 2\left[ \frac{4\pi t_{\perp}}{\hbar \omega _{c}}-\frac{\pi 
}{4}+\frac{\phi }{2}\right] \right) \right] 
{R_D^{\ast }}^{2}\Bigg\}  \nonumber 
\end{eqnarray} 
where $\phi =\arctan (a)$\ and\ $a=\hbar \omega _{c}/2\pi t_{\perp}$. 
$R_T=\left( 2\pi ^{2}k_{B}T/\hbar \omega_{c}\right) /\sinh \left( 
2\pi ^{2}k_{B}T/\hbar \omega _{c}\right)$ is the temperature damping factor 
which is equal to that in the LK theory and comes from the integration with 
the Fermi distribution function. The scattering factor $R_D^{\ast}$ 
has the same form as $R_D$ but includes a different Dingle temperature 
$T_D^{\ast}$ instead of $T_D$ as will be discussed below. 
The coefficient $\sigma _{0}$ in Eq. (\ref{Sigf}) can be estimated as $\sigma 
_{0}=e^{2}N_{LL}2t^{2}d^{2}/\pi \hbar ^{2}\omega _{c}k_{B}T_{D}$ where 
$N_{LL}$ is the Landau level degeneracy, $N_{LL}/\hbar \omega _{c}=m^{\ast 
}/2\pi \hbar ^{2}$ is the DoS at the Fermi level. 

The second term in the curly brackets of Eq. (\ref{Sigf}) describes the 
fundamental SdH oscillations modulated with the frequency 
$F_{\text {beat}}=2t_{\perp}m/e\hbar$. 
It is the last term in Eq. (\ref{Sigf}) that gives the slow oscillations 
of $\sigma_{zz}$ (hence of $R_{\perp}\propto 1/\sigma_{zz}$)
with the frequency equal to the double beat frequency, in agreement with the 
experiment. The oscillations of $F_{\text {beat}}$ as a function of the angle 
$\theta$, in accordance with the AMRO effect \cite{Wosn3,Yam} lead to 
identical oscillations of $F_{\text {slow}}$ that explains the angular 
dependence plotted in Fig.~2. It is interesting that, unlike many anomalies 
which were studied earlier and associated with a strong enhancement of the 
oscillation amplitude and harmonic contents, the present phenomenon exists 
even in the limit of a weak amplitude and constant chemical potential, i.e. 
when no substantial deviations from the standard LK description were 
expected. 

From Eq. (\ref{Sigf}) it becomes clear why the slow oscillations are 
virtually independent of temperature (Fig.~1). Indeed, they are determined 
by the interlayer transfer integral (i.e. warping of the FS) 
but not by the electron energy. Therefore, they are not affected by the 
temperature smearing of the Fermi distribution function and do not contain 
the factor $R_T$. This is why the slow oscillations, though originating from 
the cyclotron motion on the {\em same large orbits\ } on the FS cylinder as 
the fundamental SdH oscillations, persist up to relatively high temperatures 
at which the latter are completely suppressed. 

Another notable feature of the slow oscillations is that their Dingle 
temperature $T_D^{\ast }$ is different from $T_D$ entering the Dingle factor 
of the usual SdH oscillations. The usual Dingle temperature includes all 
mechanisms of the smearing of DoS oscillations. These are not only 
microscopical scattering events but also macroscopic spatial inhomogeneities 
of the sample. These inhomogeneities lead to macroscopic spatial variations 
of the electron energy in Eq. (\ref{Sn1}). Their effect is equivalent to the 
local shift of the chemical potential $\mu$. The total signal is an average 
over the entire sample and such macroscopic inhomogeneities lead to the 
damping of the magnetic quantum oscillations similar to that caused by 
temperature. Since the slow oscillations do not depend on $\mu$, they are not 
affected by this type of smearing and the corresponding Dingle temperature 
$T_D^{\ast }$ is determined by only short-range scatterers. One can therefore 
estimate relative contributions from macroscopic inhomogeneities and from 
local defects to the scattering rate by comparing $T_D$ and $T_D^{\ast }$. In 
Fig. 3 we plot the normalized amplitudes of the fundamental and slow 
oscillations as functions of inverse magnetic field. For the fundamental 
oscillations an angle $\theta = 19.3^{\circ}$ has been chosen. This angle 
corresponds to the AMRO peak, thus complications in the determination of the 
amplitude due to the beats could be avoided. Fits of the data according to 
Eq. (\ref{Sigf}) (solid lines in Fig. 3) yield the Dingle temperatures 
$T_D=(0.8\pm0.02)$ K and $T_D^{\ast}=(0.15\pm0.02)$ K. It is unlikely that 
the big difference between these values is only caused by the slightly 
different orientations. Therefore, we conclude that inhomogeneities play an 
important role in damping the SdH oscillations in the present sample. 

Although the scattering is dominated by crystal imperfections at low 
temperature, electron-electron and electron-phonon interactions must also be 
taken into account in both $R_D$ and $R_D^{\ast }$. This leads to a finite 
temperature dependence of the amplitude of the slow oscillations. A correct 
evaluation of this effect based on exact calculations of the imaginary part 
of the electron self-energy would allow to obtain additional information 
about the many-body interactions from the slow oscillations. 

Finally, we note that the interference between the oscillating relaxation 
time and interlayer Fermi velocity considered above is not the only possible 
source of the slow oscillations. There are other oscillating quantities which 
do not enter Eq. (\ref{Sn1}) directly but are contained in the relaxation 
time \cite{condmat} and can, in principle, also lead to a similar effect. 
Taking into account these additional factors will lead to a change in the 
magnitude and phase of the slow oscillations, leaving, however, their 
dependence on temperature and magnetic field essentially the same. 

In conclusion, we have shown that the slow oscillations in the Q2D metal 
$\beta $-(BEDT-TTF)$_2$IBr$_2$ originate from the warping of its cylindrical 
FS. We propose a model explaining this phenomenon as a general feature of 
clean Q2D metals developing when the cyclotron energy becomes comparable with 
the interlayer transfer energy. In particular, slow oscillations observed in 
some other layered organic metals \cite{Kar3,Wosn2,Ohmi,togo} may have the 
same origin. Since the slow oscillations are not affected by the temperature 
smearing of the Fermi distribution function, one can use them for estimating 
the warping of the FS even at relatively high temperatures at which the 
fundamental SdH oscillations are suppressed. On the other hand, the 
temperature and field dependencies of the slow oscillations can give a 
valuable information on scattering processes. 

The authors thank  A.M. Dyugaev and I. Vagner for useful discussions. The 
work was supported by the EU ICN contract HPRI-CT-1999-40013, and grants 
DFG-RFBR No. 436 RUS 113/592 and RFBR No. 00-02-17729a.

\newpage 

\begin{center} 
Figure captions 
\end{center} 

Fig. 1. Interlayer resistance (a) and oscillating part of magnetic torque (b) 
of $\beta $-(BEDT-TTF)$_2$IBr$_2$ versus magnetic field at 
$\theta\approx -15^{\circ}$. The curves at different temperatures are offset 
for clarity. 

Fig. 2. Angular dependencies of the frequency of the slow oscillations 
(a) and background resistance at $B=14$ T, $T=0.44$ K (b). Lines are guides 
for the eye. 

Fig. 3. Amplitudes of the fast (a) and slow (b) oscillations normalized to 
the background resistance plotted versus inverse magnetic field at $T=0.44$ 
K. The lines are fits to Eq. (\ref{Sigf}).


\begin{thebibliography} {99}

\bibitem{Shoe} D. Shoenberg, {\it Magnetic oscillations in metals} 
(Cambridge University Press, Cambridge, 1984). 

\bibitem{Jero} W. Kang et al., 
Phys. Rev. Lett. {\bf 62}, 2559 (1989); 
V.N. Laukhin et al.,  
Physica B {\bf 211}, 282 (1995). 

\bibitem{Harr} N. Harrison et al., 
Phys. Rev. B {\bf 54}, 9977 (1996). 

\bibitem{Wosn1} J. Wosnitza et al., 
Phys. Rev. B {\bf 61}, 7383 (2000). 

\bibitem{Balt} E. Balthes et al., 
Z. Phys. B {\bf 99}, 163 (1996); 
N. Harrison et al., 
Phys. Rev. B {\bf 58}, 10248 (1998). 

\bibitem{PhSh} P.D.Grigoriev et al., 
Phys. Rev. B {\bf 65}, 060403(R) (2002). 

\bibitem{Gv1}  V.M. Gvozdikov, Fiz. Tverd. Tela {\bf 26}, 2574 (1984) 
[Sov. Phys. Solid State {\bf 26}, 1560 (1984)]; 
T. Champel and V.P. Mineev, Phil. Magazine B {\bf 81}, 55 (2001). 

\bibitem{dHvADiff}  K. Jauregui et al., 
Phys. Rev. B  {\bf 41,} 12922 (1990); 
M.A. Itskovsky et al., 
Phys. Rev. B {\bf 61}, 14616 (2000); 
P. Grigoriev and I. Vagner, Pis'ma Zh. Eksp. Teor. Fiz. {\bf 69}, 139 
(1999) [JETP Lett. {\bf 69}, 145 (1999)]; T. Champel, Phys. Rev. B {\bf 64}, 
054407 (2001). 

\bibitem{Pavel2}  P. Grigoriev, Zh. Eksp. Teor. Fiz. {\bf 119}(6), 1257 
(2001) [JETP {\bf 92}, 1090 (2001)]. 

\bibitem{ShubDiff}  A.E. Datars and J.E. Sipe, Phys. Rev. B {\bf 51}, 4312 
(1995); V.M. Gvozdikov, Fiz. Nizk. Temp. {\bf 27}, 956 (2001) 
[Sov. J. Low Temp. Phys. {\bf 27}, 704 (2001)]. 

\bibitem{Av}  N.S. Averkiev et al., 
J. Phys.: Condens. Matter {\bf 13}, 2517 (2001). 

\bibitem{Kar1} M.V. Kartsovnik et al., 
Pis'ma Zh. Eksp. Teor. Fiz. {\bf 47}, 302 (1988) 
[Sov. Phys. JETP Lett. {\bf 47}, 363 (1988)]. 

\bibitem{Kar2} M.V. Kartsovnik et al., 
Pis'ma Zh. Eksp. Teor. Fiz. {\bf 48}, 498 (1988) 
[Sov. Phys. JETP Lett. {\bf 48}, 541 (1988)]. 

\bibitem{Kar3} M.V. Kartsovnik et al., 
Pis'ma Zh. Eksp. Teor. Fiz.  {\bf 49}, 453 (1989) [Sov. 
Phys. JETP Lett. {\bf 49}, 519 (1989)]. 

\bibitem{Wosn2} J. Wosnitza et al., 
Synth. Metals, {\bf 85}, 1479 (1997). 

\bibitem{Ohmi}  E. Ohmichi et al., 
Phys. Rev. B {\bf 57}, 7481 (1998). 

\bibitem{togo}  B. Narymbetov et al., 
Eur. Phys. J. B {\bf 5}, 179 (1998); 
T. Togonidze et al., 
Physica B {\bf 294-295} 435 (2001).

\bibitem{OMRev}  T.~Ishiguro, K.~Yamaji and G.~Saito, {\em Organic 
Superconductors}, 2nd Edition, Springer-Verlag, Berlin, 1998. 

\bibitem{Weis} H. Weiss et al., 
Phys. Rev. B {\bf 60}, R16259 (1999). 

\bibitem{Wosn3} J. Wosnitza et al., 
Synth. Metals {\bf 55-57}, 2891 (1993); 
J. Wosnitza et al., 
J. Phys. I France {\bf 6}, 1597 (1996). 

\bibitem{Kar4} M.V. Kartsovnik et al., 
J. Phys. I (France) {\bf 2}, 89 (1992). 

\bibitem{comment0} The azimuthal orientation of the field projection onto 
the $ab$-plane was different for the measurements corresponding to Figs. 1 
and 2. This is why the value $F_{\text {slow}}=42$ T corresponding to Fig. 1a 
does not exactly fit to the curve in Fig. 2a. 

\bibitem{Yam}  K. Yamaji, J. Phys. Soc. Jpn. {\bf 58}, 1520 (1989). 

\bibitem{AMRO} R. Yagi et al., 
J. Phys. Soc. Jpn. {\bf 59}, 3069 (1990); 
V.G. Peschansky et al., 
J. Phys. I (France) {\bf 1}, 1469 (1991); 
Y. Kurihara, J. Phys. Soc. Jpn. {\bf 61}, 975 (1992). 

\bibitem{Mah}  G. Mahan, {\it Many-Particle Physics}, 2nd ed. (Plenum Press, 
New York, 1990). 

\bibitem{condmat} P.D. Grigoriev et al., 
cond-mat/0108352 (unpublished). 

\end{thebibliography}
\end{document}